\documentclass[10pt, conference]{IEEEtran}
\IEEEoverridecommandlockouts
\usepackage{cite}
\usepackage{amsmath,amssymb,amsfonts}
\usepackage{algorithmic}
\usepackage{graphicx}
\usepackage{textcomp}
\usepackage{xcolor}
\newcommand{\etal}{\textit{et al. }}
\usepackage{url}
\usepackage{subcaption}
\def\BibTeX{{\rm B\kern-.05em{\sc i\kern-.025em b}\kern-.08em
    T\kern-.1667em\lower.7ex\hbox{E}\kern-.125emX}}
\newcommand{\TDCattack}{attack scheduler}
\newcommand{\ourtit}{DeepStrike}
\usepackage{url}

\begin{document}

\title{DeepStrike: Remotely-Guided Fault Injection Attacks on DNN Accelerator in Cloud-FPGA
\thanks{This work was supported in part by the National Science foundation under grants SaTC-1929300, CNS-1916762, and SaTC-2043183.}}
\author{\IEEEauthorblockN{Yukui Luo$^{\diamond}$, Cheng Gongye$^{\diamond}$, Yunsi Fei, and Xiaolin Xu}\IEEEauthorblockA{{\small{(${\diamond}$ indicates equal contribution)}}\\{Department of Electrical and Computer Engineering, Northeastern University, Boston, MA, USA}}}

\maketitle
\IEEEpubidadjcol
\begin{abstract}
As Field-programmable gate arrays (FPGAs) are widely adopted in clouds to accelerate Deep Neural Networks (DNN), such virtualization environments have posed many new security issues. This work investigates the integrity of DNN FPGA accelerators in clouds. It proposes DeepStrike, a remotely-guided attack based on power glitching fault injections targeting DNN execution.  We characterize the vulnerabilities of different DNN layers against fault injections on FPGAs and leverage time-to-digital converter (TDC) sensors to precisely control the timing of fault injections. Experimental results show that our proposed attack can successfully disrupt the FPGA DSP kernel and misclassify the target victim DNN application.
\end{abstract}


\section{Introduction}
The recent advancement of deep learning has made it a powerful tool in solving various challenging problems with superb performance. 
Many real-world applications have high throughput requirements and a stringent power consumption budget for the deep neural network (DNN) engines. 
Hardware-based DNN accelerators have been proposed and deployed on different computing platforms, including graphic processing units (GPUs), application-specific integrated circuits (ASICs), and field-programmable gate arrays (FPGAs). FPGA has shown unique advantages among different types of platforms, offering higher design and implementation flexibility than ASICs and higher power efficiency than GPUs \cite{FPGAbeatGPU}. 
Leading cloud service providers such as Amazon~\cite{AmazonEC3:online} and Microsoft~\cite{DeployML91:online} have integrated powerful FPGAs in their cloud servers, enabling machine learning as a service (MLaaS). The commercialization of MLaaS has facilitated deep learning in various compute-intensive applications, including medical diagnosis assistance~\cite{AIMachin80:online} and risk and fraud management~\cite{GlobalMa7:online}. 




To increase the resource utilization and reduce the cost of cloud services, many recent works have enabled cloud-FPGA to be shared by multiple users, i.e., independent tenants utilize an FPGA chip in their allocated time or concurrently \cite{provelengios2020power,zha2020virtualizing}. However, such a co-tenancy usage model of cloud-FPGA also poses new security issues and creates new attack surfaces. 
In \cite{Krautter_Gnad_Tahoori_2018}, Krautter \etal showed successful fault injection attacks on AES running on an FPGA, in which the adversary utilizes a periodically enabled power-hungry circuit to disrupt the FPGA power distribution network (PDN). As a result, the victim AES circuit, sharing the same PDN, generates transient computation errors which lead to faulty ciphertext outputs. Differential fault analysis (DFA) then utilizes the faulty outputs to retrieve the secret key. 

The wide deployment of DNNs on cloud-FPGA has rendered DNN engines a new vulnerable victim to potential security attacks.
There have been some prior fault injection attacks on DNNs, targeting either a microcontroller using laser beam~\cite{deeplaser} or DRAMs with software row-hamming~\cite{255272}. Several recent 
attacks on DNN FPGA implementations use hardware fault injection 
such as memory collisions~\cite{8844478}, clock glitch~\cite{rakin2020deep}, and weight loading
perturbation~\cite{9218577}.

This paper presents \ourtit, a novel fault attack on DNN accelerators in cloud-FPGA with power-glitching fault injections. 
Unlike the existing fault injection attacks requiring full knowledge of the DNN model implementation, \ourtit~ deduces the execution details of the victim DNN model through side-channel analysis. Towards this goal, we propose to leverage an on-chip delay-sensor built with time-to-digital converters (TDC)\cite{gnad2017voltage, zick2013sensing}. The delay sensor can identify the execution of different DNN layers with high temporal resolution. 
Informed by such execution details of the victim DNN model from TDC, the adversary can remotely guide and launch the fault injections with fine timing control. 
We propose a novel power striker to induce glitches on the power distribution network of cloud-FPGA, disrupting the victim DNN execution with fault injection. Unlike other commonly used power-hungry circuits, the proposed circuit scheme can pass the design rule checking (DRC), making it a viable design choice.
We characterize the fault sensitivities of different types of DNN layers. With such knowledge, the fault injections are guided to target the most vulnerable DNN layers, making the end-to-end attack more efficient and stealthy.
We demonstrate the effectiveness of the proposed attack with LeNet-5 architecture implementation on Xilinx PYNQ-Z1 FPGA with MNIST dataset.


\if false. 
The main contributions of this paper are as follows:
\begin{itemize}
\item We propose a remotely-guided fault injection strategy that does not require the implementation details of the victim DNN model. To the best of our knowledge, this is the first work that uses side-channel leakage to guide fault injections on DNN accelerators in cloud-FPGA.

\item We propose a novel power striker to induce glitches on the power distribution network of cloud-FPGA, disrupting the victim DNN execution with timing violations. Unlike other commonly used power-hungry circuits, the proposed circuit scheme can pass the design rule checking (DRC) checking, making it a viable design choice.

\item We characterize the vulnerabilities of different DNN layers against fault injections, and target our fault strikes at the most vulnerable DNN layers. 

\item We demonstrate the effectiveness of the proposed attack with LeNet-5 architecture implementation on Xilinx PYNQ-Z1 FPGA with the MNIST dataset.
\end{itemize}
\fi 


The rest of the paper is organized as follows. Section~\ref{sec:background} presents the background and related work. Section~\ref{sec:deepstrike} illustrates the proposed \ourtit\ with important components and attack procedures. Section~\ref{sec:case_study} describes our end-to-end attack experiments and analyzes the results. Section~\ref{sec:discussion} concludes this paper and discusses the future work. 

\section{Background and Related work}\label{sec:background}
\subsection{Threat model}\label{sec:threat_model}
This work adopts a common threat model of cloud-FPGA used by many other related works \cite{ramesh2018fpga,giechaskiel2018leaky,yazdanshenas2019costs,khawaja2018sharing}. It can be summarized as follows: 1) Enabled by FPGA virtualization, multiple users co-reside on an FPGA chip and there is no physical interaction between the circuits of different users, and these circuits can execute simultaneously \cite{luo2019hill}.
2) All users of the same cloud-FPGA chip share certain hardware resources like the power distribution network (PDN). 3) The two circuit applications running on the cloud-FPGA are from a benign user and an adversary, respectively, i.e., a DNN accelerator is the victim and a malicious circuit aims to breach the integrity of the victim execution. 
Additionally, we consider two more, strict but realistic, conditions in our threat model: 4) The adversary does not have implementation details of the target DNN model, nor access to the DNN input and output (i.e., a black-box attack). 5) The design rule checking of modern cloud-FPGA does not allow the implementation of combinational loop circuit, such as a ring-oscillator (RO).

\subsection{Power Distribution Network of FPGAs}
Sharing the FPGA hardware resources between different users make it possible for the adversary to interfere with other co-located benign users. 
Among the shared resources, the PDN of a cloud-FPGA becomes a new attack surface with all the users sharing it. 
Recently several attack methods targeting PDNs have been presented. 
Confidentiality of the victim application can be breached by passive side-channel leakage. 
Various on-chip sensors, such as TDC and RO, have been designed to infer the behavior of the victim FPGA users. For example, the transient power trace of a victim RSA encryption engine is sensed by RO-based power sensors for off-line key retrieval~\cite{zhao2018fpga}. In another prior work~\cite{schellenberg2018inside}, TDC is used to capture the transient voltage fluctuations of the victim application for side-channel attacks.
The TDC-based delay-sensor is also constructively used as a sensor for defending the FPGA against power side-channel attacks~\cite{gnad2018checking}. Moreover, the integrity of the victim application on cloud-FPGA can also be compromised by active fault injections by malicious users, as detailed next. 
 


\subsection{Related Work}
A few recent works have also explored the security of DNN implementations on FPGAs. In \cite{8844478}, Alam \etal proposed to attack the DNN model through memory collision. Specifically , they inject faults to the DNN model by writing complementary data to both ports of a memory cell. 
Liu \etal \cite{9218577} utilized clock glitches to introduce timing violations to the DNN accelerators on FPGA so as to cause misclassification. 
In \cite{fault_sneak}, Zhao \etal simulated the performance of DNN models under fault injection attacks. Specially, they randomly choose and flip certain parameters of the DNN model and test corresponding model accuracy. 

Although these existing attacks demonstrate effectiveness in reducing the inference accuracy of DNN models, several 
important drawbacks have limited their practical applicability: 1) Most work \cite{8844478}\cite{9218577}\cite{fault_sneak} adopt a white-box attack, in which the adversary has full knowledge of the victim DNN model as well as implementation details (e.g., the memory location of DNN parameters), which is impractical. 2) Some attack scheme  \cite{fault_sneak} is only validated with simulation, which may not be applicable to real FPGA DNN implementations. 



\section{DeepStrike Attack}\label{sec:deepstrike}

\begin{figure*}[tb!]
    \centering
    \includegraphics[width=0.9\linewidth]{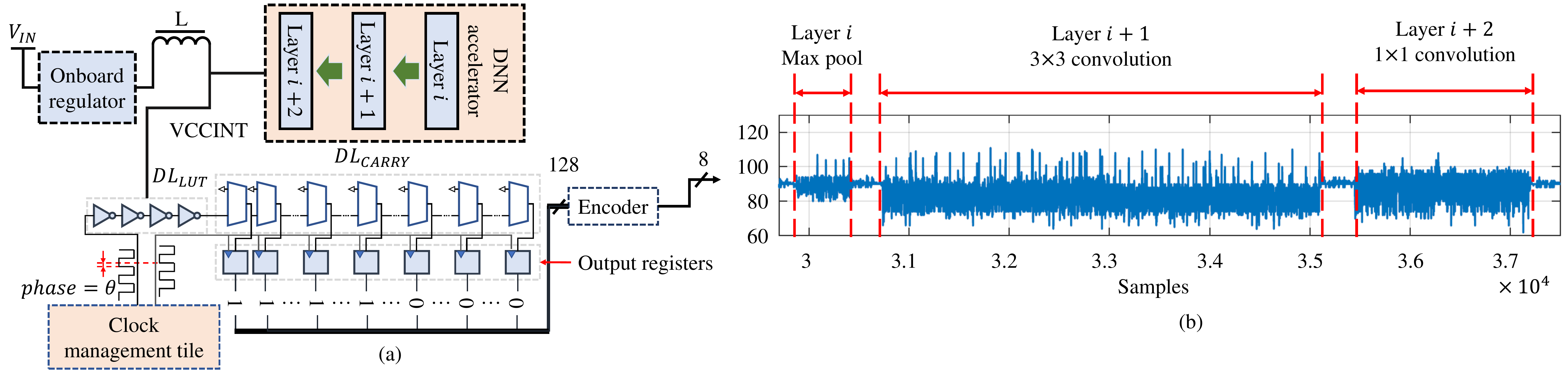}
    \caption{(a) The proposed TDC-based delay-sensor and victim DNN accelerator 
    sharing the power distribution network on an FPGA. (b) Voltage fluctuation associated with three DNN layers' execution collected by TDC-based delay-sensor.}
    \label{Fig:TDC_sch}
\end{figure*}

\subsection{Attack Overview}
\subsubsection{The Victim}
We target DNN FPGA accelerators as the victim, 
leveraging parallel high-performance processing engines (PEs). These engines are typically implemented by digital signal processing (DSP) slices, the dedicated hardware units on modern FPGAs for acceleration. For example, in DNN accelerators, DSPs are mainly utilized to speed up multiplications and summations. Additionally, the DSP slice is also one of the most used hardware components by state-of-the-art Xilinx Deep Learning Processor Unit (DPU)\cite{pg338}. 

\subsubsection{The Attacker}\label{sec:atk_step}
The proposed attack mainly consists of two salient parts, namely \textit{Attack scheduler} and \textit{Power striker}. The \TDCattack\ is important and responsible for 1) Monitoring and profiling the victim DNN model execution through side-channel leakage (e.g., transient voltage fluctuation) and 2) Activating the power striker at critical timing points. Directed by the well-informed attack scheduler, the power striker will inject faults to the execution of specific DNN layers, in a targeted fashion. Specially, we utilize the TDC-based delay-sensor and a novel power-wasting circuit to construct the attack scheduler and power striker, respectively. 
More detailed design schemes of these two parts are presented below.

\if false 
Our attacks is consist of three steps. The first step to obtain the execution information of the victim. We achieved this by mobilizing the TDC-based delay-sensor to gather the side-channel information leakage. The details of this step is discussed in section \ref{subsec:TDC}.

The second step is analyzing the information gathered in the previous step. The adversary would need to inspect and waveform and identify the time interval of the execution of each layer. Then one could test which layer is more susceptible to the attack.  This step is explained in section \ref{sec:attacking_scheduler}

The last step is launching the attack using power glitch fault injection. In section \ref{subsec:power_glitch} we described our fault injection capability and analyzed the fault characteristics.
\fi 

\subsection{Attack Scheduler}\label{sec:tdc}

The schematic of the \TDCattack\ is illustrated in Fig. \ref{Fig:TDC_sch}(a), which mainly consists of a TDC-based delay-sensor, a clock management tile, and an encoder. Taking FPGA implementation as an example, the TDC circuit is composed of two elements: $DL_{LUT}$, a look-up-table (LUT) based delay-line, and $DL_{CARRY}$, a carry-chain built with MUX and D flip-flop. The length of $DL_{LUT}$ determines the resolution of the TDC-based delay-sensor, and the $DL_{CARRY}$ can scale the output range of the TDC. During the operation of TDC-based delay-sensor, two clock signals of the same frequency will be generated by the clock management tile. One clock drives the $DL_{LUT}$, and the other clock is for sampling the registers connected to the carry-chain outputs. There exists a phase difference $\theta$ between these two clocks, which is used for calibrating the readout. The direct output of TDC is a binary vector generated by the carry chain, which consists of different numbers of consecutive ``1s" and ``0s'' determined by the voltage/delay. The encoder can convert these direct outputs of registers into a binary code, i.e., from 128-bit to 8-bit unsigned int value (to count the number of ``1"s in the 128 bits), as the sensor readout.  

Since the propagation delay of the two clock signals through the delay-lines is closely impacted by the transient voltage level, the TDC sensor readout becomes an indicator of the real-time voltage. In other words, when the FPGA is executing applications, the voltage will fluctuate and the readings of the sensor can depict the voltage profile. As illustrated in Fig. \ref{Fig:TDC_sch}, while the TDC-based delay-sensor shares PDN with another circuit application (e.g., DNN model), its readout can be used to profile the voltage fluctuation caused by the execution of that application. In practical usage, the driving clock frequency ($F_{dr}$) and the length of $DL_{LUT}$ ($L_{LUT} $)and $DL_{CARRY}$ ($L_{CARRY}$) should be carefully designed to avoid counting errors.

\if false
the ideal output method, a consecutive ``1" output starts at the least significant bit (LSB) output, and the rest output are all zeros. We count the number of consecutive ``1" as the output to represent the internal components’ propagation delay status, and we set the unit as “bin” \cite{zick2013sensing}. 
While the power attack applies to the corresponding FPGA, the bin will reduce. 
\fi 

A primary challenge for a remotely-guided fault attack on a multi-user FPGA is that the attacker does not have knowledge of the model execution. 
To mitigate this issue, we propose to use the TDC-based delay-sensor to profile and infer the target DNN model execution. In our preliminary study, we sequentially execute three layers of a DNN model: a max-pooling layer, a convolutional layer with a $3 \times 3$ kernel, and a convolution layer with a $1 \times 1$ kernel. Meanwhile, the TDC readout is collected in parallel. The specific configuration of the TDC-based delay-sensor for this victim is $F_{dr} = 200MHz$, $L_{LUT} = 4$, $L_{CARRY} = 128$, and we calibrate $\theta$ to get approximate 90 consecutive "1" outputs when the FPGA works under a nominal voltage. Fig. \ref{Fig:TDC_sch}(b) gives a tracing example for the tested DNN execution, which shows that the sensor readouts clearly present different patterns for executions of different DNN layers. We also notice clear ``stalls'' between different layer executions (the readout stays around 90), and the fluctuation during convolutional layers' execution is much larger than that of the max-pooling layer. Therefore, we conclude that the side-channel leakage of the victim DNN model execution can be used to build a library of sensor readout patterns for different types of DNN layers at different sizes for future attack use. 


\if false
which implements the layer $i$ to the layer $i+2$. The layer $i$ is a 
\fi



\subsection{Power Striker}\label{sec:attacking_scheduler}
\begin{figure}[tb!]
    \centering
    \includegraphics[width=0.5\linewidth]{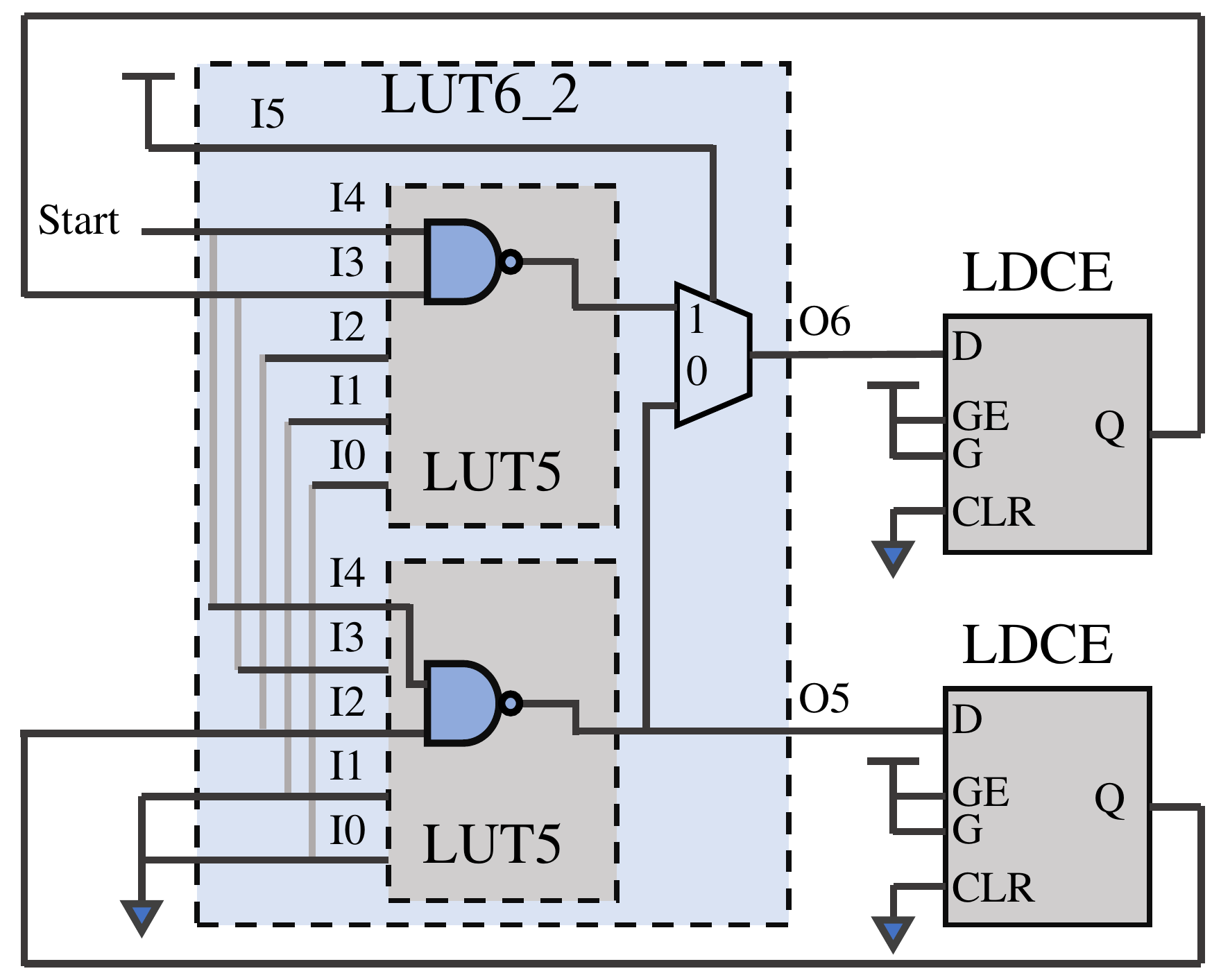}
    \caption{A controllable power striker design scheme.}
    \label{Fig:PW_circuit}
\end{figure}
Another important component of the proposed \ourtit\ attack is the power striker. It is a malicious controllable power-wasting circuit used for aggressively overloading the shared PDN, incurring well-timed voltage glitches.  
The design requirement for the power striker is even when the malicious circuit is activated for a short period (e.g., a few clock cycles), it draws a significant amount of power, creating an immediate voltage drop on the shared PDN. As a result, the voltage drop increases the signal propagation time in FPGA components that share the same PDN, inducing timing violations and computation or data loading faults \cite{Krautter_Gnad_Tahoori_2018}. 
Previous works mainly utilized LUT-based combinational loops (e.g., RO) to construct such malicious circuits \cite{la2020fpgadefender,provelengios2020power}. Although those circuit schemes are effective, they trigger design rule checking (DRC) warnings and are commonly banned by security and privacy-sensitive cloud-FPGAs \cite{8730895}. 

We develop a circuit scheme that can pass the DRC checking by inserting data latches in the combination loop, for our \textit{power striker}. Fig. \ref{Fig:PW_circuit} depicts the basic circuit cell, which utilizes a two-output LUT ($LUT6\_2$) with two latch registers (LDCE). 
When enabled (\texttt{Start}=1), the $LUT6\_2$ is configured as two parallel inverters, with their outputs, $O6$ and $O5$, connected to two LDCEs, respectively, to form two oscillating loops. Compared with the combinational loop, this method increases the loop's length and utilizes one LUT for two self-oscillating loops. As a result, the proposed circuit scheme can provide higher attack efficiency with less hardware overhead. Moreover, it can pass the DRC checking. 
An adversary of cloud-FPGA can apply a large number of such power striker cells, and use the \texttt{Start} signal to control the duration of their activation. 

\subsection{Attack Scheduler and Power Striker Integration}
As described in Sec. \ref{sec:threat_model}, our threat model assumes that a practical attacker may not have any knowledge about the victim DNN model’s parameters. Thus, a fine-tuning attack on the specific weight or pixel computing is impossible. Instead, \ourtit\ targets at activating the power striker multiple times, starting at a guided moment by the attack schedule, i.e., during the execution of a particular DNN layer. 
As illustrated in Sec. \ref{sec:tdc}, the TDC-based delay-sensor can be used to track and characterize the execution of target DNN. Once enough characteristics (e.g., time duration, TDC readout, etc.) for each distinct DNN layer are gathered, we can build a profile to assist with scheduling the activation of the power striker. Practically, the profiling procedure can be accomplished during the normal target DNN model, i.e., classifying different input images.

\begin{figure}[tb!]
    \centering
    \includegraphics[width=\linewidth]{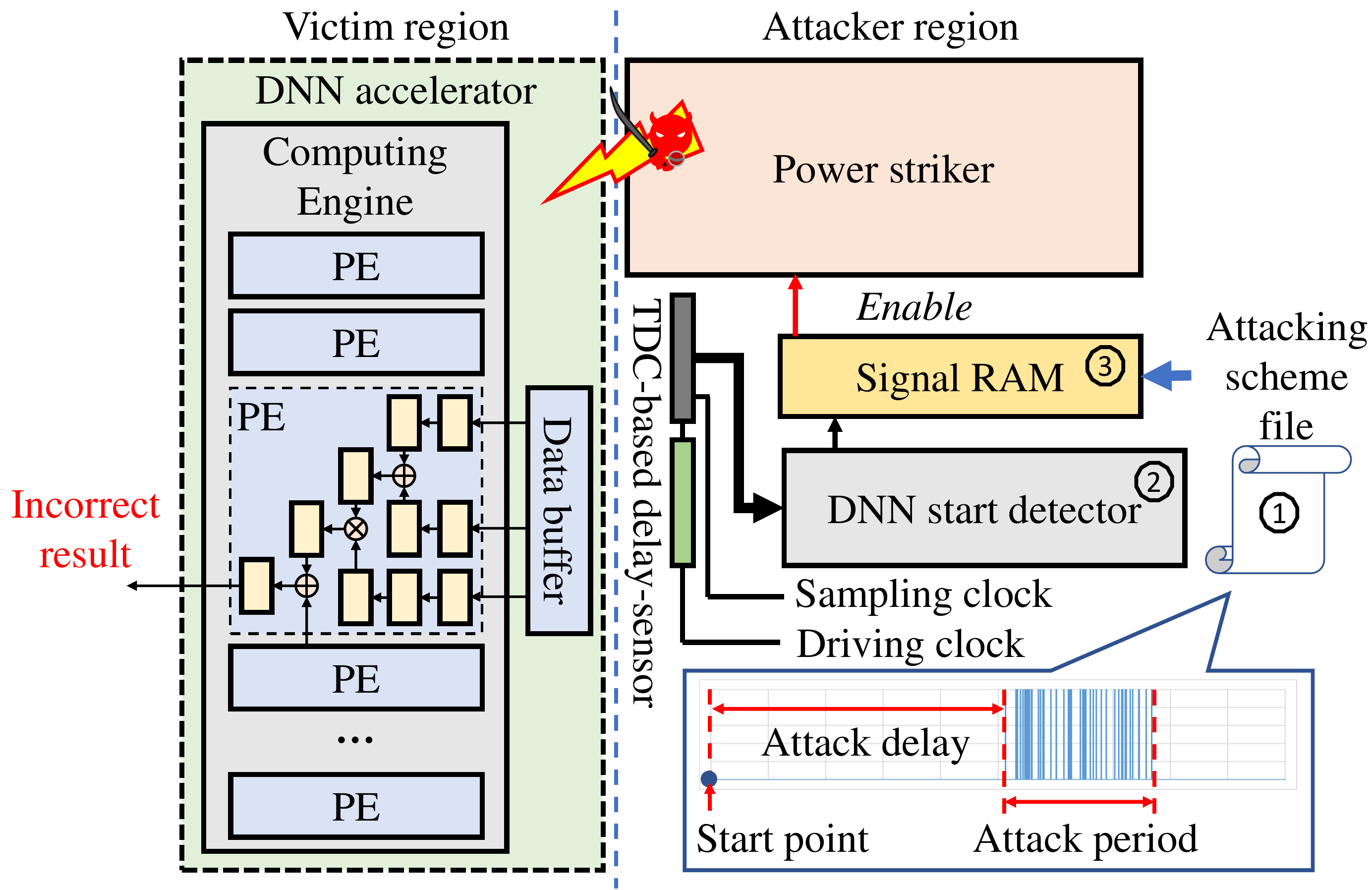}
    \caption{Integrative schematic of \ourtit.}
    \label{Fig:ATK_scheduler}
\end{figure}

Fig. \ref{Fig:ATK_scheduler} shows the integrative schematic of the attack scheduler and power striker, including some other auxiliary circuits/components like the 
\textit{DNN start detector} and \textit{signal RAM}. The design schemes and functionalities of these components are as follows.

\subsubsection{DNN start detector}
\begin{figure}[htpb]
    \centering
    \includegraphics[width=1\linewidth]{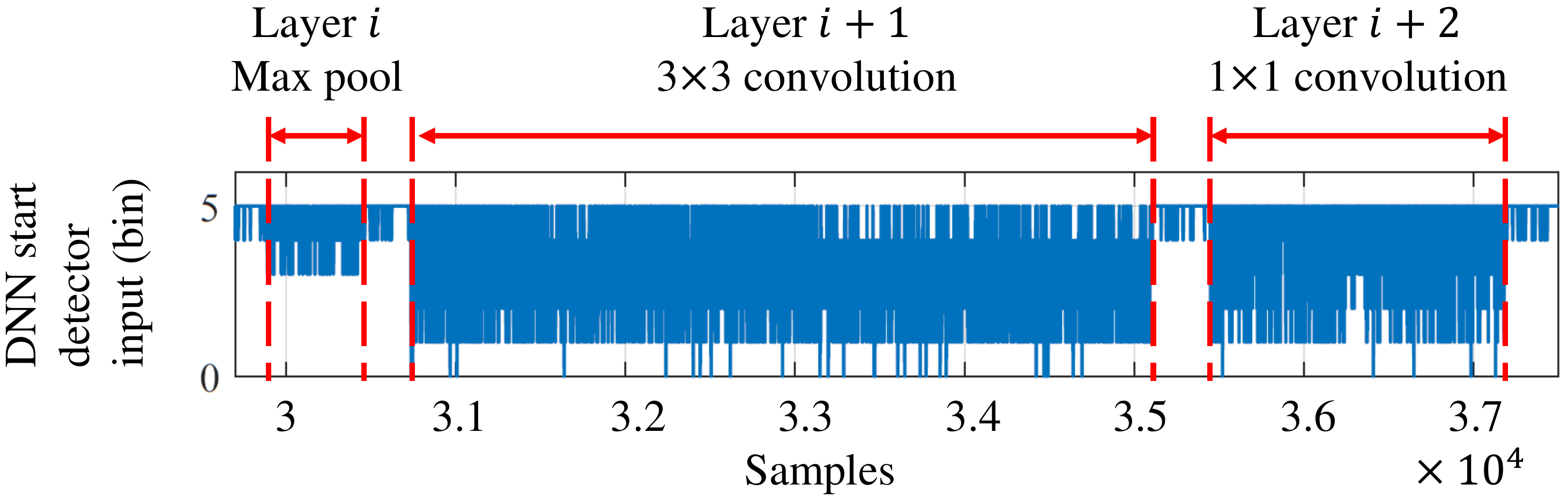}
    \caption{Input of the DNN start detector. }
    \label{Fig:DNN_detect}
\end{figure}

From Fig. \ref{Fig:TDC_sch}(b), we can observe that there always exist small voltage fluctuations (i.e., the ``stall'' zones) on the FPGA PDN even when the DNN models are not being executed. These small voltage fluctuations, although can be detected by the TDC-based delay-sensor, cannot be used to guide the proposed attacks. Thus, to filter out the impact of these small voltage fluctuations, we need to purify the voltage fluctuation sensed by the TDC sensor. To realize this, we build the \textit{DNN start detector} with a finite-state machine (FSM), with its inputs connected to outputs of the TDC-based delay-sensor. We partition the 128-bit TDC output into five zones, and select 1-bit from each zone as the input of the \textit{DNN start detector}. Leveraging such voltage fluctuation purification, we apply the \textit{DNN start detector} to detect the DNN model (The same DNN model we used in Fig. \ref{Fig:TDC_sch})
execution, and the results are shown in Fig. \ref{Fig:DNN_detect}. Compared to the results by the TDC-based delay-sensor shown in Fig. \ref{Fig:TDC_sch}(b)), the purified voltage fluctuation can provide  more accurate and controllable guidance to start the \ourtit\ attack. For example, when the \textit{DNN start detector} gets an input Hamming weight (HW) equals to 3, indicating the first layer - \texttt{MaxPool} just starts, we set up a ``start point'' for our attack scheduler. 

\subsubsection{Signal RAM}
To make the proposed attack configurable, we develop another component \textit{signal RAM} with the on-chip BRAM, which is used to store the \textit{attacking scheme file}. The attacking scheme file mainly includes three parameters: \textit{attack delay}, \textit{attack period}, and the \textit{number of attacks}. Specifically, these parameters are denoted as binary vectors and each bit represents the action of \ourtit\ during a separate clock cycle. 
We use ``1'' to enable and ``0'' to disable the power striker, respectively. Therefore, the parameter \textit{number of attacks} can be configured by using different 1/0 composition. Additionally, to control the time duration (i.e., clock cycles) elapsed before enabling a power strike, we define \textit{attack delay}, which is represented by a series of ``0s''. With the \textit{signal RAM} (i.e., on-chip BRAM) being read at a specific clock frequency $f_{sRAM}$, the duration of \textit{attack delay} is jointly determined by the number of ``0s'' in it and $f_{sRAM}$. For example, a \textit{attack delay} consisting $N$ ``0s" will pause \ourtit\ for $N$ clock cycle, with time duration of $\frac{N}{f_{sRAM}}$. Similarly, the duration of \textit{attack period} is configured in this way with consecutive ``1"s. 




In summary, the proposed \ourtit\ attack can be accomplished in three steps: 1) Profiling the voltage fluctuation associated with the victim DNN accelerator execution to make a corresponding attack plan, i.e., determining these different parameters like \textit{number of attacks}, \textit{attack delay}, etc., store these parameters in the \textit{signal RAM}; 2) Using \textit{DNN start detector} to sense the execution of victim DNN accelerator; 3) Launching \ourtit\ following the pre-scheduled attack strategy in \textit{signal RAM}. 
We would like to highlight that with the proposed attack scheme, the attacker have high flexibility to load different attack strategies at run-time, i.e., dynamically target at different DNN layers. 

\section{Experimental setup and validation results}\label{sec:case_study}
In this section, we present an end-to-end attack experiment on a PYNQ FPGA evaluation kit, which is an open-source project that integrates the Linux system with the Xilinx FPGA. 
Here we apply the Xilinx PYNQ-Z1 FPGA board to build a prototype of the cloud-FPGA. Without loss of generality, in our experimental validation, we choose an open-source DNN accelerator engine \cite{yolov2github} and train a LeNet-5 neural network \cite{lecun2015lenet,Wang_FCCM18} with the MNIST dataset~\cite{MNIST_data}. 
In our threat model, the hypervisor in the virtualized cloud-FPGA will compile and combine applications of all the tenants (including the attacker's malicious circuits and the victim’s DNN inference), generate an unified bitstream and deploy it on one FPGA device~\cite{zha2020virtualizing}. Note that although the tenants co-locate on the FPGA, they do not share hardware including the I/O bus, BRAM, and clock sources. In our experiment, the adversary connects to this prototyped cloud-FPGA from the UART serial port, with which 
the adversary can gather on-chip side-channel leakage from the TDC-based delay-sensor and dynamically configure the the \textit{attacking scheme file}.

\begin{figure}[tb!]
    \centering
    \includegraphics[width=0.9\linewidth]{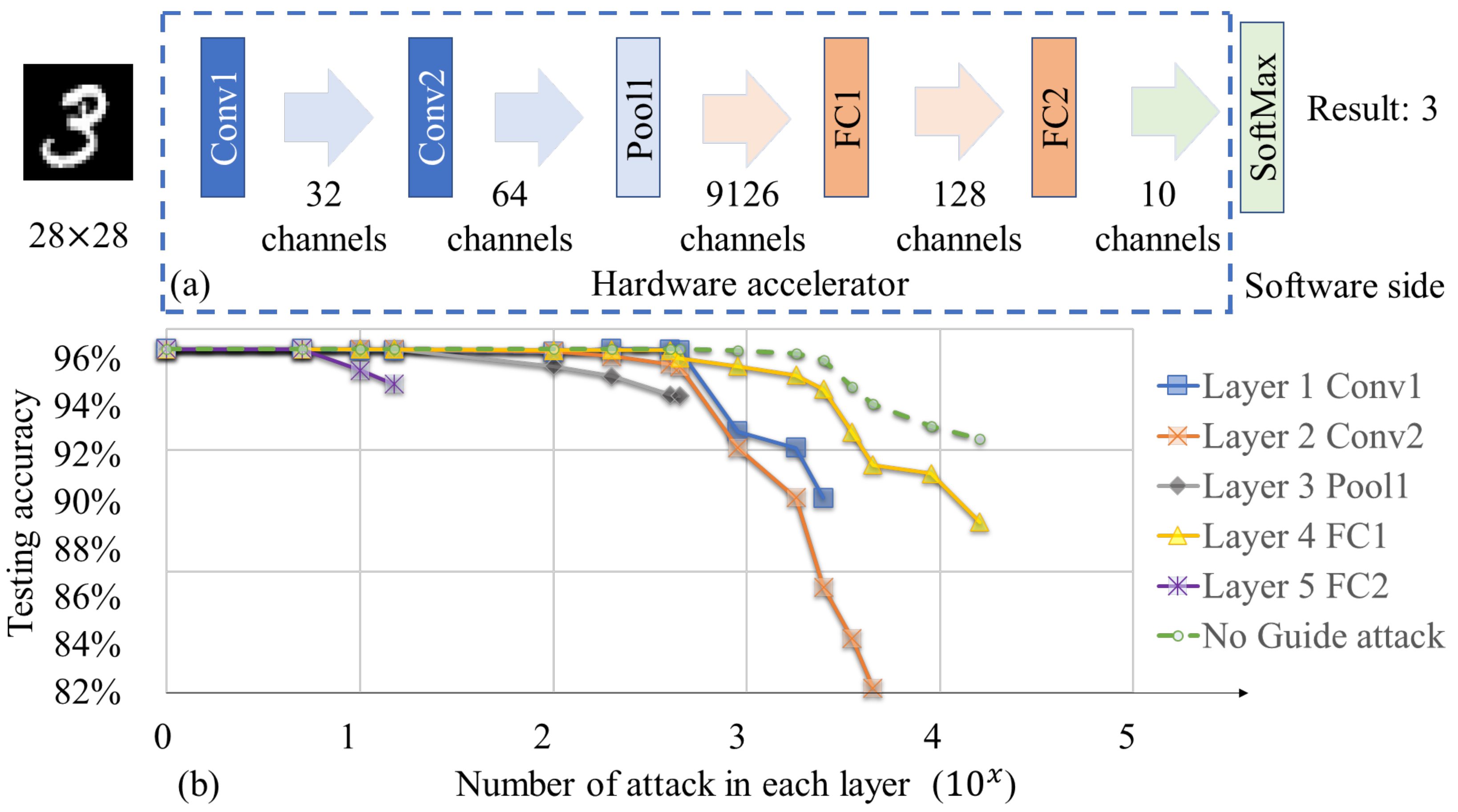}
    \caption{Case study: apply DeepStrike on MNIST application.}
    \label{Fig:MNIST_Result}
\end{figure}

The pre-trained LeNet-5 model on the MNIST dataset is deployed on the prototype cloud-FPGA.  
The data type of the model is fix-point 8-bit value, with 3-bits for the integer and the rest for the mantissa representation. The MNIST dataset includes $60,000$ training samples and $10,000$ testing samples. Our un-tampered model achieves a testing accuracy of $96.17\%$ on the FPGA. The architecture of LeNet-5 is shown in Fig. \ref{Fig:MNIST_Result}(a), which consists of two convolutional layers for feature extraction (\texttt{Conv1} and  \texttt{Conv2}), one pooling layer for downsampling (\texttt{Pool1}), and two fully connected layers (\texttt{FC1} and \texttt{FC2}) for classification. The output of the \texttt{FC2} is a vector of 10 prediction scores, which go through a SoftMax layer to pick the class with the largest score as the prediction. 
Note as we use the unsigned fixed-point quantization method, the activation function we use in this case study is the hyperbolic tangent (\texttt{tanh}).

We target each layer separately and apply a series of fault injections while the corresponding acceleration kernel is executing, guided by the \textit{attack scheduler}. The power striker circuit consumes $15.03\%$ logic slices, and each power glitching strike lasts for $10 ns$. 
We observe the inference accuracy to evaluate the end-to-end effect of fault injections on different layers. 
Fig. \ref{Fig:MNIST_Result} (b) shows that the testing accuracy drops as the number of power strikes increases. 
Note that due to the different execution length of different layers, the maximum number of strikes on different layer also varies. As observed in the results, \texttt{CONV2} is the most fault-sensitive layer, and the maximum accuracy drop reaches $14\%$ when $4500$ strikes are applied.  
Additionally, we provide the results of non-TDC guiding attacks as our baseline, which is the top curve, where the fault injections happen randomly along with the model execution. We conclude that our proposed TDC guiding DeepStrike fault attack is much more efficient than the blind attack while applying the same attack intensity. 

Moreover, the experimental results show that the vulnerability to the power glitching fault injections of each layer depends on the layer's type, the layer's size, and its execution time. As \texttt{CONV2} is larger than \texttt{CONV1} and takes longer to execute, more fault injection strikes can be applied onto \texttt{CONV2} and result in the largest testing accuracy reduction. \texttt{FC1} takes the longest time to execute. However, it is a fully connected layer and only adds $k \times k$ prior multiplication results to generate one pixel in a feature map. Convolution layers contain more complex multiplications. We find that these most vulnerable layers (e.g., \texttt{CONV2} and \texttt{FC1} ) are implemented with DSP slices. One reason that DSP slice-based DNN layers are more vulnerable lies in the design rules. To increase the performance of the DNN accelerators, the designers usually adopt double-data-rate while using DSP, enabling doubled running speed of the DSP slices compared to other components. This design choice, although makes the DSP slices faster, also renders it more vulnerable to fault injection attacks due to the tighter timing constraints.


\subsection{Fault Characterization of DSP Slices under Power Strikes}\label{sec:DSP}
\begin{figure}[tb!]
    \centering
    \begin{subfigure}{0.38\linewidth}
        \centering
        \includegraphics[width=1\linewidth]{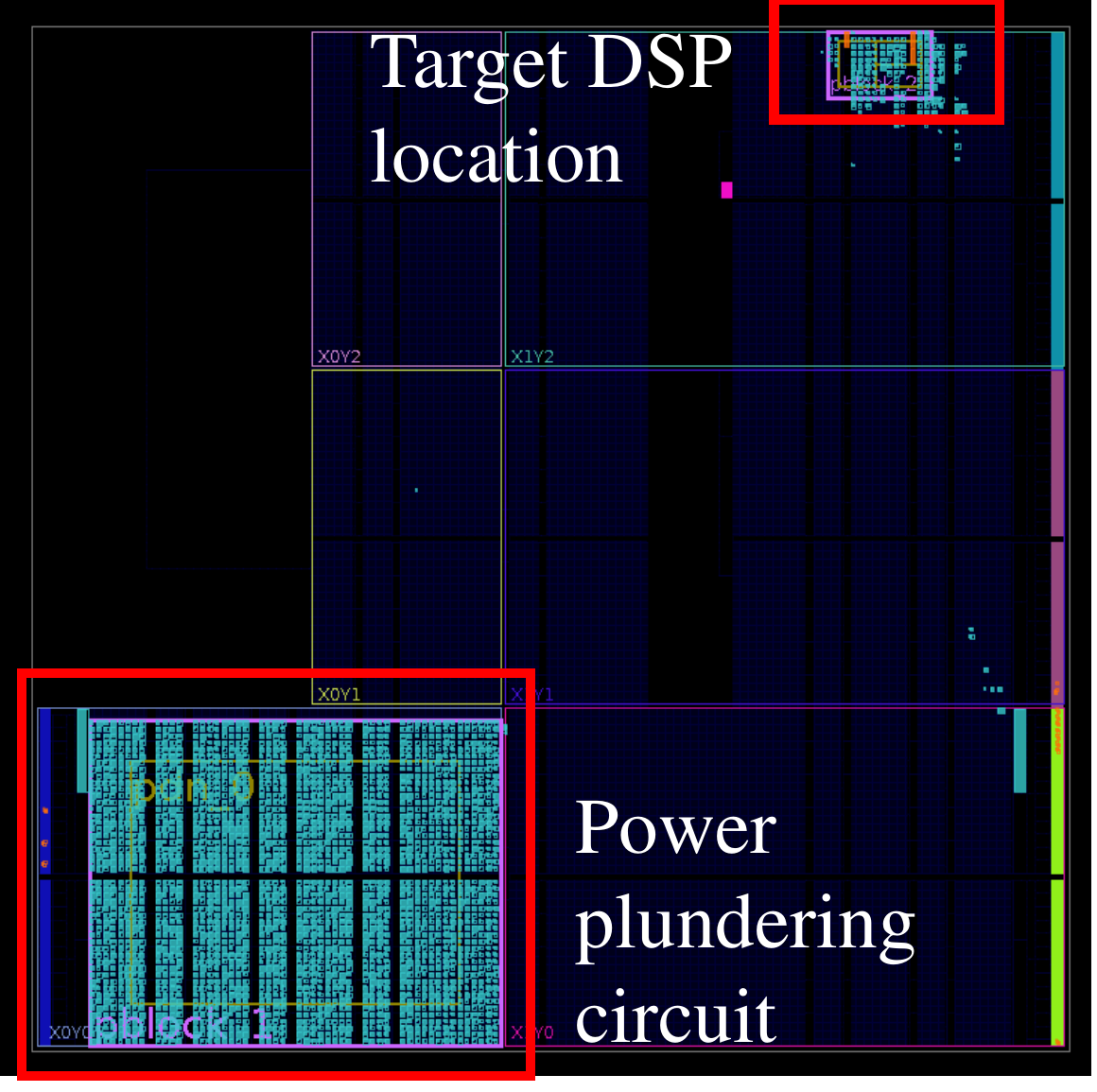}
        \caption{The Xilinx FPGA (XC7Z020) layout for DSP fault injection test.}
        \label{Fig:layout_DSP}
    \end{subfigure}
    ~
    \begin{subfigure}{0.5\linewidth}
        \centering
        \includegraphics[width=1\linewidth]{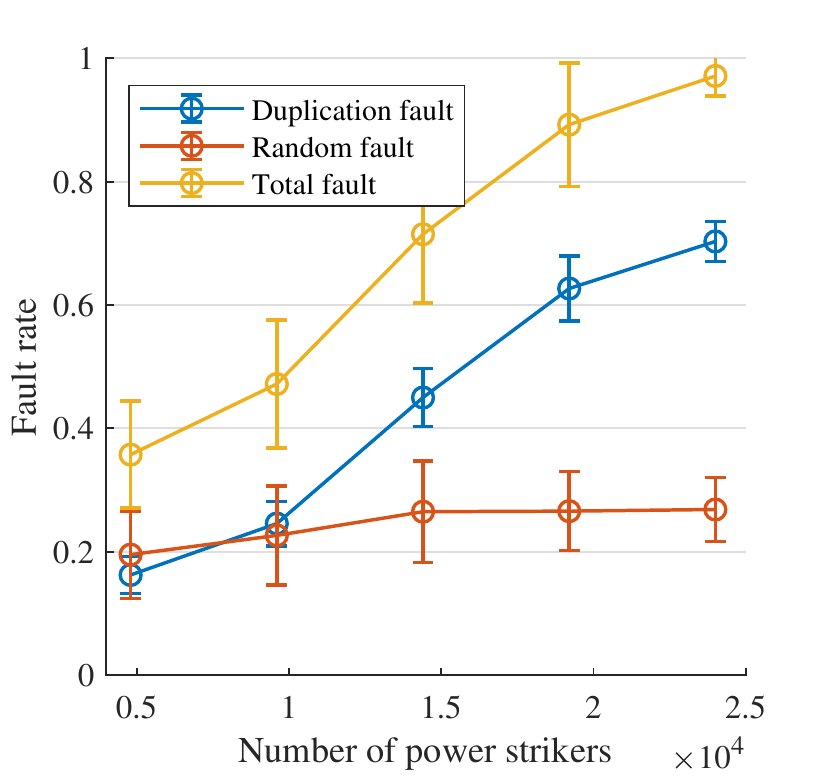}
        \caption{Duplication fault  rate and random fault rate of double data rate DSP slices with different numbers of power striker cells.}
        \label{Fig:DSP_result}
    \end{subfigure}
    \caption{DSP fault injection test: configurations and results.}
\label{Fig:DSP_exp}
\end{figure}

We designed experiments to investigate the faults in DSP slices caused by power glitching strikes. The layout of the attack is shown in Fig. \ref{Fig:layout_DSP}. We put the victim circuit far from the attacker circuit to minimize the influence of temperature changes, which sharing the PDN. The DSP slices are configured to add two inputs and multiply with the third input, which is the configuration for convolution computation \footnote{The fully connected layers are usually implemented on DSP slices with these configurations as they could be treated as a special case of convolution}. Since the DSP slices do not have a result-ready signal, we  designed a circuit that fetches the result of the DSPs after five clock cycles. This circuit works correctly and the timing analysis of the FPGA mapping tool does not complain about violations of timing constraints. 

We fed the DSP slices 10,000 randomly generated inputs and launched the power striker circuit for one clock cycle at the same time we enabled the DSP slices. According to our experiment, we only need to enable the attack for one cycle to induce fault in a single DSP computation operation. Enabling the power striker circuit longer will work as well but it may increase the temperature of the FPGA chip or even crash it.



We observed two types of faults from the experimental results, namely 1) Duplication fault, where the DSP output is the correct result of the previous input. In this case, the DSP computation simply takes more cycles to complete and cannot produce the correct result in time; and 2) Random faults, where the faulty output does not have obvious patterns. In Fig. \ref{Fig:DSP_result}, we demonstrated both types of faults, in which the x-axis is the number of power striker cells, and the y-axis denotes the fault rate, number of faults divided by the total number of experiments. The experimental result shows that we can control the fault injection intensity by adjusting the number of power striker cells. For example, the total fault rate\footnote{The total fault rate is the sum of duplication fault rate and random fault rate. When launching the proposed attack in practice, much fewer power striker cells is needed because other victim components also consume power, further reducing the voltage of the PDN and strengthening fault injection.} is nearly 100\% with 24,000 power strike cells. 

In conclusion, the power glitching fault injection results in random faults or duplication faults in the DSP slices. With duplication faults, the correct product appears in the next clock cycle, and can be absorbed by more serial summations, mitigating the adverse impact of stale results in \texttt{FC} layers. Also convolutional layers involve much more multiplications, possibly experiencing more random faults and making them more  vulnerable. These experimental results well explain that \texttt{FC1} achieves much less accuracy reduction than \texttt{CONV2} under the same number of fault injection strikes. 


\section{Conclusion and Future work}\label{sec:discussion}
We demonstrate \ourtit, a remotely-guided power glitching fault injection attack targeting DNN accelerators in cloud-FPGA. Different from other attacks that require implementation details of the victim DNN model, 
\ourtit leverages the voltage fluctuations associated with the on-chip DNN model execution as a side-channel information, to launch well-scheduled fault injection attacks. We prototyped an experimental cloud-FPGA on a PYNQ FPGA development board, and conducted end-to-end attacks to validate the effectiveness of the proposed attack scheme on a LeNet-5 neural network trained with the MNIST dataset. The experimental result demonstrates that the proposed attack scheme can significantly lower the inference accuracy. We also investigate the possible reasons why different DNN model layers show different resilience against power glitching fault injections. 

In future work, we plan to extend 
the proposed attack scheme to more complicated execution environments, e.g., more than three tenants on the FPGA, which may be representative multi-user scenarios for cloud-FPGA.   
We will also consider more DNN architectures, and experiment with commercial cloud-FPGAs. 



\bibliographystyle{IEEEtran}
\bibliography{bibliography/main.bib}
\end{document}